\documentclass[preprint]{vgtc}               

\graphicspath{{figures/}{pictures/}{images/}{./}} 

\usepackage{times}                     

\usepackage{tabu}                      
\usepackage{booktabs}                  
\usepackage{lipsum}                    
\usepackage{mwe}                       

\usepackage{mathptmx}                  
\usepackage{amsfonts}
\usepackage{graphicx}
\usepackage{enumitem}
\usepackage{soul}

\newcommand{\revision}[1]{#1}

\onlineid{1198}

\vgtccategory{Research}

\vgtcinsertpkg

\title{Toward a Logic of Generalization about Visualization as a Decision Aid}

\author{Alex Kale\thanks{e-mail: kalea@uchicago.edu}\\ %
\scriptsize University of Chicago %
\vspace{-5mm}
}

\abstract{
    Visualization as a discipline often grapples with generalization by reasoning about how study results on the efficacy of a tool in one context might apply to another context.
    This work offers an account of the logic of generalization in visualization research and argues that it struggles in particular with applications of visualization as a decision aid.
    We use decision theory to define the dimensions on which decision problems can vary, and we present an analysis of heterogeneity in scenarios where visualization supports decision-making.
    Our findings identify utility as a focal and under-examined concept in visualization research on decision-making, demonstrating how the visualization community's logic of generalization might benefit from using decision theory as a lens for understanding context variation.
} 


\begin{document}


\firstsection{Introduction}

\maketitle


Visualization research aims to make generalizable claims about the effectiveness of various data representations for supporting decision-making.
In this pursuit, we confront a trade-off between the internal and external validity of our study designs.
Although controlled experiments (e.g.,~\cite{Fernandes2018, Wesslen2022-myopic}) enable us to measure the relative effectiveness of visualization designs with precision, it can be unclear when such findings are expected to be applicable to contexts that are different from the conditions of the study~\cite{Hullman2025-underspecified}.
Conversely, design studies~\cite{Sedlmair2012-trenches} offer deep insights into domain context, albeit with less precision about any particular operationalization of effectiveness.
Both approaches contribute to visualization research in important ways, yet we struggle to synthesize them into a cohesive framework of design knowledge.
We argue that these epistemic challenges are especially pronounced for applications of visualization to decision-making, where \textit{we lack a sufficient logic of generalization for reasoning about context distance}---i.e., differences between the conditions of a study vs. the conditions of a scenario where we might like to apply the results of that study, which make us question the transportability and applicability of study results.

The epistemology of visualization research spans a continuum between optimism and caution about generalization.  
The traditional model of visualization ``effectiveness''~\cite{Mackinlay1986APT} frames visualizations as vehicles for data communication, such that user performance on all sorts of downstream tasks can be explained by how well as visualization supports ``atomic''~\cite{Cleveland1984} perceptual judgments involved in decoding data from an image.
Research in this paradigm pursues pragmatic goals such as building recommendation systems (e.g.,~\cite{Mackinlay1986APT})
and making design recommendations intended for general use.
However, it also commits to a techo-solutionist~\cite{Rosner2018} pursuit of universal 
knowledge,
which may be unrealistic insofar as it assumes we can isolate the atomic tasks involved in decoding a visualization in a manner invariant to context.
We dub this logical move toward generalization the \textit{optimistic atomicity assumption}.

Recent research 
\revision{has}
critiqued and offered alternatives to the epistemic status quo 
\revision{of optimistic atomicity}.
For example, visualization research has made strides to condition on context in research on personalization (e.g.,~\cite{oscar2017-personal_viz}), individual differences (e.g.,~\cite{Ottley2015-individual}), and the needs of people with disabilities (e.g.,~\cite{lundgard2021accessible}).
Far on the cautious side of the epistemic continuum are problem-focused design studies~\cite{Sedlmair2012-trenches} and related qualitative methods, which make a principled commitment to design context-specific abstractions and offer rich descriptions of the context in which they were developed (e.g.,~\cite{McCurdy2019-implicitError}).
Such attention to context provides critical input to our logic of generalization, however, there is a risk that researchers and practitioners engaging with a study will struggle to reason about the conditions under which an abstraction or technique would be applicable or effective, especially absent definition of how findings are thought to be context-dependent.
We dub this logical move of rigorously documenting context while remaining uncommitted about its role in generalization, assuming that others will sort it out on a case-by-case basis, the \textit{cautious indeterminacy assumption}.

\begin{figure}[t]
    \centering
    \includegraphics[width=\linewidth]{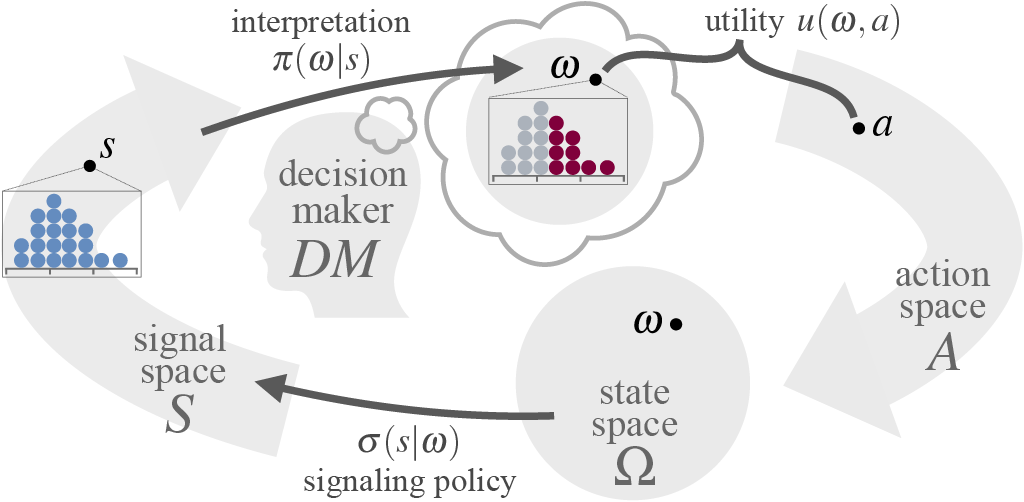}
    \caption{Decision theory describes problems in terms of how a decision-maker might optimally use the information provided (left) to choose an action in line with a utility function (right).}
    \label{fig:decision-theory}
\end{figure}

Neither of these logical moves provides an adequate basis for generalization about visualization as a decision aid because neither offers a formal way of accounting for how context matters for decision-making.
Both 
nonetheless agree that it is important \textit{how visualizations are used}, even while they lack a shared theoretical vocabulary for relating effectiveness to use in a 
situated context. 
Decision theory~\cite{savage1972foundations} offers such a vocabulary. 
As illustrated in Figure~\ref{fig:decision-theory}, it links how visualizations are generated and interpreted 
(left) with how visualizations are used to inform actions (right). 
Building on recent work by Hullman et al.~\cite{Hullman2025-underspecified, Wu2024-rational}, we explore decision theory as a framework for reasoning about how decision problems differ structurally in ways that would be expected to influence the design goals and efficacy of decision aids.
\looseness=-1

We contribute this provocation for visualization researchers with the intention of promoting decision theory as a tool for reasoning more systematically about context variation in decision problems.
First, we present a primer on decision theory for visualization research, focused on how this formal framework identifies the dimensions along which decision problems can vary.
Then, we use decision theory to systematically analyze forms of heterogeneity in examples of decision problems from the visualization literature and application domains, showing how variations in decision context such as the relationships among decision-makers can influence design goals for visualization. 
We discuss the epistemological implications of heterogeneity in decision problems for visualization research, namely the pivotal role of utility in defining decision context and ways that the field might better account for context variation in constructing design knowledge about decision support.

\section{Heterogeneity in Decision Problems}
To characterize forms of heterogeneity in scenarios where visualizations might be deployed as decision aids, we draw on decision theory~\cite{savage1972foundations} for a formal definition of dimensions of decision problems. 
\revision{Our conceptual analysis builds on pioneering mathematical accounts of what can be learned from visualization experiments by Wu et al.}~\cite{Wu2024-rational}\revision{ and Hullman et al.}~\cite{Hullman2025-underspecified}.
This \revision{approach} enables a systematic account of possible variations in decision contexts and why these variations matter when making design recommendations.

\subsection{Visualization ($\sigma$) According to Decision Theory}
Figure~\ref{fig:decision-theory} 
provides a decision-theoretic~\cite{savage1972foundations} account of how visualization interpretation factors into decision-making.
The diagram depicts how a decision-maker can choose an optimal action in \textbf{two main steps}: (1) she uses a given visualization $s$ to form an interpretation $\pi$ of the state of the world $\omega$ (left); and (2) she uses this inferred state $\omega$ along with a utility function $u$ to select among possible actions $a \in A$ (right).
Canonically, the decision-maker's task is to choose the optimal action such as to maximize the function:
\begin{equation}
\label{eq:1}
\mathop{argmax}\limits_{a \in A} \mathop{\Sigma}\limits_{\omega \in \Omega} u(\omega, a) \pi(\omega|s)
\end{equation}
In this formalism, visualization techniques are mappings between a space of possible states $\omega \in \Omega$ and a space of possible signals $s \in S$, which decision theorists refer to as a \textit{signaling policy} $\sigma: \Omega \rightarrow \Delta S$, where $\Delta S$ is a discrete probability distribution over possible signal realizations $s \in S$. 
\revision{This captures the intuition that} a visualization technique \revision{produces} a specific signal realization $s$ (i.e., a chart with a particular appearance) \revision{from the state $\omega$ through a stochastic process, which we can alternatively express as} a conditional probability $\sigma(s|\omega)$.
\revision{Normatively, $\pi$ represents a Bayesian posterior (see}~\cite{Hullman2025-underspecified,Wu2024-rational}\revision{), but in practice users may interpret visualizations heuristically (e.g.,}~\cite{Kale2021}\revision{).}
At the bottom left of Figure~\ref{fig:decision-theory}, a visualization designer proposes a technique $\sigma$ mapping between the state of the world $\omega$ and a visualization $s$ (blue dotplot) sent to the decision-maker $DM$.
At the top of Figure~\ref{fig:decision-theory}, $DM$ interprets $\pi$ the realized visualization $s$ (counting the purple dots inside the thought bubble) in an attempt to invert the mapping $\sigma$ and infer a task-relevant state $\omega$.
\looseness=-1

Visualization research mostly focuses on Step 1, the construction $\sigma$ of signals $s$ and how decision-makers interpret them to understand or predict $\pi$ the state of the world $\omega$.
Usually, we are hoping to generalize about the efficacy of interpretation for different possible visualization techniques (i.e., signaling policies), in the spirit of optimistic atomicity.
However, recent work~\cite{Jardine2020, Kale2021} suggests that visualization users don't all use the same cognitive strategies for interpretation, and may not even be self-consistent in their strategies.

Step 2, valuation of actions $a \in A$ in light of a utility function $u$, receives relatively little attention from visualization research (e.g., see~\cite{Dimara2018-evaluating, Fernandes2018, Kale2025-PB, Verma2023-allocation} for exceptions).
Instead, most studies of what Hullman et al.~\cite{Hullman2025-underspecified} call ``fully specified'' decision problems endow a utility function through the use of instructions and incentives (e.g.,~\cite{Hofman2020-effects, Kale2021, Oral2024-decoupling, Sarma2024-odds, Wesslen2022-myopic, Zhang2024-strategic}), rather than investigating users' actual preferences.
More often, visualization research leaves the utility function open-ended and describes participant's revealed preferences through their behavior (e.g.,~\cite{Levy2021-MedicalDM}), or it measures interpretation $\pi$ (e.g., using Likert responses~\cite{Padilla2017-hurricane})
and then speculates about how visualization might support downstream decisions.

We show that \textit{utility functions are implicated in many important variations of decision problems}, moderating both the desirability of actions $a \in A$ and the states $\omega$ that are task-relevant and thus the target of interpretation $\pi$ for the decision-maker.
Without accounting for the interplay of dimensions in a decision task (Sections 2.2--2.5), the logic of generalization in visualization research collapses important distinctions between classes of decision problems.
\looseness=-1

\subsection{Variations in Decision-Maker ($DM$) Relationships}
Consider 
\revision{how}
people can work together to make data-driven decisions 
\revision{with}
visualizations. 
Settings with more than one decision-maker will require a different formal model than \autoref{eq:1}.

\textbf{The Unitary Decision-Maker.} 
Visualization research tends to emphasize settings where an individual decision-maker makes a definitive choice independent of the actions of others.
We typically use this kind of task setup to study the individual cognitive process of a decision-maker interpreting data.
\textit{For example, consider Kay et al.'s studies~\cite{Fernandes2018, Kay2016-WhenIsMyBus} of transit decisions based on visualized predictions of bus arrival times.}
Participants judge cumulative probability of catching their bus $\omega$ from a univariate distribution of predicted bus arrival times $s$ (e.g., similar to the interpretation depicted in Figure~\ref{fig:decision-theory}) in order to decide when $a$ to be at the bus stop.
Notably, the second study in this series~\cite{Fernandes2018} surveyed transit riders about their preferences $u$ to determine realistic incentives for a subsequent experiment, a relatively unique effort to seed a normative evaluation with a real-world utility function.
If we interpret these studies using the \textit{optimistic atomicity} assumption, we might claim that the efficacy of interpretation $\pi$ for the visualizations tested $\sigma$ would generalize to other use cases involving  cumulative probability judgments on a continuous univariate distribution $\omega$, whether it's bus arrival times or otherwise, in a way that we hope is separable from the utility function $u$ and cognitive challenges of evaluating it. 
However, from the perspective of decision theory, the interpretation $\pi$ and utility $u$ are inexorably linked in ways that makes such generalizations hard to justify, although some have tried to distinguish the cognitive processes of decoding from decision-making~\cite{Kale2021, Oral2024-decoupling}.

\textbf{A Distributed Network of Decision-Makers.}
In some contexts, multiple decision-makers act within a network, and their decisions may impact each other in complex ways.
\textit{For example, consider the task in Zhang et al.~\cite{Zhang2024-strategic} where crowdworkers played the role of rideshare drivers choosing which zone to travel to for their next pickup.}
Because there are multiple decision-makers $DM_i \in \{DM_1, ... DM_N\}$, settings like this entail a different objective function, such as maximizing
wellfare over all participants, \looseness=-1
\begin{equation}
\label{eq:2}
\mathop{argmax}\limits_{a_1, ... a_{N} \in A} \mathop{\Sigma}\limits_{i = 1}^{N} \mathop{\Sigma}\limits_{\omega \in \Omega} u(\omega, a_i) \pi_i(\omega|s)
\end{equation}
where $a_i$ is the action chosen by $DM_i$\revision{, and $\pi_i$ is} $DM_i$\revision{'s interpretation of a visualization} $s$.
Zhang et al.~\cite{Zhang2024-strategic} investigated the design of visualizations of traffic congestion $\sigma$ that were shared with all participants, assuming that each participant attempts to maximize their bonus using the same scoring rule $u$.
However, it's easy to imagine task variations, such as persuasion campaigns on social media (e.g.,~\cite{Lee2021-misinfo}), where each decision-maker $DM_i$ might see personalized visualizations $\sigma_i$ and make choices based on their individual values $u_i$. 
Thus, designing for multiple decision-makers introduces considerable complexity around how information signals $s$ and utilities $u$ are distributed within networks. 

\begin{figure}[t]
  \centering
  \includegraphics[width=\linewidth]{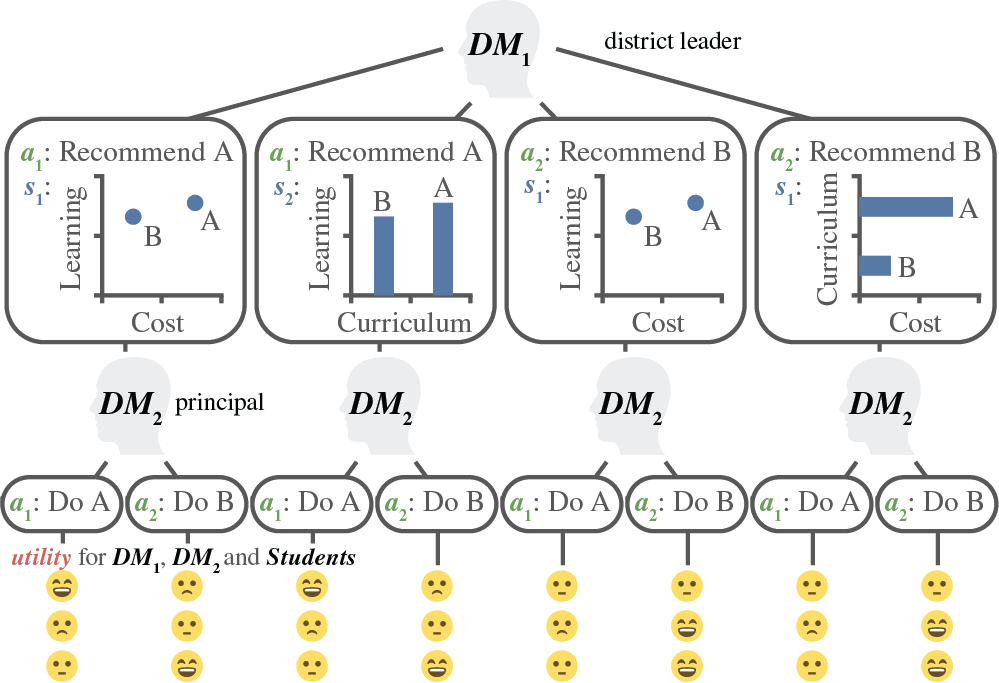}
  \caption{Sequential decision-making within an organization, where early $DM$s pass actions $a$ and signals $s$ to subsequent $DM$s.}
  \label{fig:game}
\end{figure}

\textbf{A Hierarchy of Decision-Makers.}
In organizational settings, multiple decision-makers tend to have hierarchical relationships, and these scaffold the flow of information $s$ and actions $a$.
\textit{For example, consider the context of decision-making in education, as described by Coburn et al.~\cite{Coburn2020}, where district leaders make policy recommendations which are subsequently interpreted by principals and implemented by teachers.}
\revision{A formal treatment of this decision problem requires game theory}~\cite{savage1972foundations}\revision{---e.g.,}
Figure~\ref{fig:game} offers a simplified demonstration of possible dynamics in a setting with two $DM$s at different levels of an organization.
First, a school leader $DM_1$ recommends either curriculum A $a_1$ or B $a_2$ and shares evidence about these curricula, either $s_1$ or $s_2$, with a principal $DM_2$.
Then, $DM_2$ chooses to enact either curriculum A $a_1$ or B $a_2$ in their school.
Notice how the utilities $u$ for $DM_1$, $DM_2$, and $\mathit{Students}$ all depend on \revision{the} actions taken by decision-makers at each layer of the organization.
The role of evidence $s_1$ or $s_2$ shared by $DM_1$ is to persuade~\cite{Kamenica2011-BayesPersuasion} $DM_2$ by influencing their interpretation of curriculum quality $\pi(\omega|s)$.
The design objective for visualization depends on whether $DM_1$ wants to \textit{adversarially} maximize their own utility $u$ by encouraging the curriculum $a_1$ with slightly higher learning vs. \textit{collaboratively} maximize overall wellfare (i.e., max sum of all $u$) by encouraging the curriculum $a_2$ with much lower cost.

\subsection{Variations in Action Space ($A$)}
Consider the various kinds of actions that might be available to a decision-maker, represented by $A$ in a model like \autoref{eq:1}.

\textbf{Interventions.}
Empirical evaluations of decision-making with visualization tend to focus on situations where the decision-maker $DM$ chooses a single action $a$ whose impact a visualization $s$ should help them predict.
\textit{For example, consider the tasks in Kale et al.~\cite{Kale2021}, Oral et al.~\cite{Oral2024-decoupling}, and Hofman et al.~\cite{Hofman2020-effects}, which asked crowdworkers to use visualizations $s$ to evaluate the impact $\omega$ of deploying particular interventions $a$.}
Kale et al. ask participants whether or not they will purchase an intervention $\{a, \neg a\} = A$, and Oral et al. extend this to five alternatives $\{a_1, ... a_5\} = A$, whereas Hofman et al. ask participants how much they are willing to pay $a$ within a range of prices $A$.
Despite the fact that each of these studies evaluate intervention decisions, we should be careful about directly comparing their results because (i) decision-makers are offered different action spaces $A$, (ii) different visualization techniques $\sigma$ are interpreted $\pi$ with different states of the world $\omega$ in mind, and (iii) actions $a$ and states $\omega$ are valued differently $u$.

\textbf{Plans and Strategies.}
Often, decisions are complex operational changes that entail a set of actions taken together.
In such cases, the action space $A$ can be harder to define, and the utility $u(\omega, a)$ may be harder to reason about, whether because 
plans $a$ are harder to implement than interventions or because of ambiguity about what states $\omega$ favor a particular plan.
\textit{For example, consider again the transit decisions task in Fernandes et al.~\cite{Fernandes2018}.}
This study simplifies the action space $A$ by boiling down an entire commute to the decision of when to leave, however, even with this simplification, it is difficult to interpret the uncertainty about bus arrival times $\pi(\omega|s)$ in term of utility $u(\omega, a)$, beyond the intuition that it's better to leave early than to be late.
To support decision-making as strategic planning, we may 
\revision{benefit from (i) visualizing}
utilities $u$ more directly as cost-benefit analyses (e.g.,~\cite{Kale2025-PB, Verma2023-allocation}), 
\revision{(ii)}
diagramming how actions flow together within systems~\cite{brumar2024-typology} 
\revision{and (iii) applying game-theoretic analysis}~\cite{savage1972foundations} \revision{to visualization design objectives (e.g., Figure}~\ref{fig:game}\revision{).}

\textbf{Doing Nothing.}
Although visualization research tends not to consider ``do nothing'' as a possible action~\cite{Correll2015-nothing}, this can be wise if a decision-maker has limited information or could be penalized for premature action.
Doing nothing also aligns with the well-established operational decision-making strategy of 
acknowledging sources of uncertainty without necessarily acting upon them~\cite{Lipshitz1997}.
In use cases where we value caution, we should explore using visualization in ways that inspire reflection, humility, and patience.
\looseness=-1

\subsection{Variations in Utility Functions ($u$)}
Decision theory posits that $DM$ has a utility function $u(\omega, a)$ that they optimize in order to make a choice of action $a \in A$ based on the apparent state of the world $\omega$.
Consider how the decision-maker's utility function influences the role of visualization.

\textbf{Well-Defined Utility Functions.}
Visualization research tends to assume that people are responsive to incentives like scoring rules and associated monetary bonuses.
\textit{For example, consider the task in Sarma et al.~\cite{Sarma2024-odds}, which asks participants to report insights from multiple comparisons subject to incentives that reward true positives/negatives and penalize false positives.}
The incentives suggest a desired tolerance for spurious insights $u$, enabling an investigation of whether people correct for multiple comparisons as much as they should when reporting findings from exploratory data analysis.
Hullman et al.~\cite{Hullman2025-underspecified} suggest that this reflects a sort of \textit{best case scenario} for studying decision aids, where we know the utility function $u$ exactly, assume $DM$ has perfectly internalized it, and translate the task-relevant states of the world into signals $\sigma: \Omega \rightarrow \Delta S$ in ways that highlight values of $\omega$ that correspond to extrema in utility $u$. 

\textbf{Ambiguous Utility.}
More often, we offer visualizations as decision support in situations where we could not so easily compute the utility-optimizing action because people may be uncertain about the costs and benefits associated with actions.
\textit{For example, consider the task in Levy et al.~\cite{Levy2021-MedicalDM}, which asks participants to label medical images for the purpose of cancer diagnosis without giving a clear set of incentives to specify the stakes of errors}.
In such cases, it may be realistic to assume that the stakes $u$ of medical diagnosis vary on a case-by-case basis and that participating medical practitioners interpret visualizations with these stakes in mind according to their professional standards of ethics.
However, this still poses a challenge for generalization because when utility $u$ is unclear, $DM$s will tend to select an action $a \in A$ in line with their ``cognitive default'' or prior~\cite{Enke2023, Hullman2025-underspecified}.
Under the assumption of \textit{cautious indeterminacy} (see Section 1), we tend not to systematically distinguish the impacts of utility from those of prior knowledge or other influences on user behavior, and thus we struggle to anticipate what would make a visualization effective in related applications.
Decision theory suggests we might fill this gap in our epistemology by classifying decision scenarios based on their dimensions such as the utility functions of decision-makers and stakeholders.

\textbf{Collective and Adversarial Decision-Making.}
If there are multiple decision-makers $DM_i \in \{DM_1, ... DM_N\}$ in a network or hierarchy (see above), they may each have different utility functions $u_i \in \{u_1, ... u_N\}$.
Each $DM_i$ may arrive at different actions $a_i$ based on their individual utility functions, even if they interpret visualizations the same $\pi(\omega|s)$.
\textit{For example, consider the use case in Kale et al.~\cite{Kale2025-PB}, where city residents vote to allocate funding to municipal projects through participatory budgeting.}
Residents' utility functions $u_i$ may reflect shared priorities, such that they agree on the value of specific projects (i.e., possible future states $\omega$), or residents may disagree in ways that make their advocacy and voting behavior more adversarial (cf. Figure~\ref{fig:game}).
In such situations, arguments about values $u_i$ become at least as important as facts about projects $\omega$ because $DM$s are likely to engage in motivated reasoning. 
The affordances of visualization for deliberation become important, both for convincing people to change their values $u_i$ and for persuading~\cite{Kamenica2011-BayesPersuasion} them to interpret information differently $\pi(\omega|s)$.

\subsection{Variations in the Role of Information Providers}
Consider the various communication environments in which \textit{information providers}, who we refer to as `designers' in the visualization community, might share data as a visualization. 
In a formalism like \autoref{eq:1}, the provider $P$ might choose visualizations $\sigma$ such as to persuade~\cite{Kamenica2011-BayesPersuasion} the decision-maker $DM$ to take the action $a$ that optimizes the provider's utility function $u_P$, similar to how we describe scenarios where multiple $DM$s influence each other (Figure~\ref{fig:game}).

\textbf{Open Information.}
In the most open possible relationship between the information provider $P$ and the decision-maker $DM$, both $P$ and $DM$ have access to the same data source.
This enables $DM$, in principle, to check the visualizations $s$ provided by $P$ to ensure that they reflect the state of the world $\omega$.
\textit{For example, consider the typical use case for visual analytics tools~\cite{illuminating3}, a data scientist $P$ summarizing data for their boss $DM$ to support some business decision.}
The data scientist's utility function $u_P$ is aligned with their boss's $u_{DM}$ because both their employment relationship and the openness of information naturally align the data scientist's incentives with telling the truth about the data such as to build and maintain trust.
Telling the truth here means choosing a visualization $\sigma$ so that signal realizations $s$ accurately correspond to task-relevant states $\omega$.
In visualization research, we tend to study truth-promoting situations like this, possibly out of a sense of professional ethics.

\textbf{Privileged Access.}
However, it is often the case that an information provider $P$ chooses a visualization technique $\sigma$ for an audience $DM$ who cannot access the data directly. 
Under this setup, it is easier for $P$ to pursue a goal $u_P$ that is not aligned with the decision-maker's utility $u_{DM}$.
\textit{For example, consider how internet service providers (ISP) disclose data about broadband access to consumers and government agencies in ways that make it difficult to detect suspected fairness issues~\cite{Sharma2024-equity}.}
If a government agency $DM$'s goal $u_{DM}$ is to regulate the ISP $P$ in ways that require them to provide equitable broadband access, then the ISP's utility $u_P$ will be maximized by presenting data in ways $\sigma$ that makes such regulation difficult or impossible.
The gap between utility functions $u_P$ and $u_{DM}$ for various states $\omega$ can help us recognize what visualizations $(\sigma, S)$ might be most and least effective for deception.
Formally operationalizing deception in this way might provide a helpful path toward defining enforceable standards against deceptive uses of visualization.

\textbf{Competition Among Information Providers.}
In many settings, there are multiple information providers $P_i \in \{P_1, ... P_N\}$ in competition with each other, each with their own utility function $u_{P_i}$.
Any information marketplace meets this description.
\textit{For example, consider how political groups use social media to spread misinformation, as studied with visualizations by Lee et al.~\cite{Lee2021-misinfo}.}
Different political groups $P_i$ show different messages and visualizations $\sigma$ hoping to induce a change in the decision-maker's interpretation of the state of the world $\pi(\omega|s)$.
Generally, a truthful visualization will provide a mapping $\sigma: \Omega \rightarrow \Delta S$ capable of expressing every task-relevant state $\omega$ with minimal information loss.
This definition helps us distinguish formally between use cases we would consider pro-social and those we would consider unethical.

\section{Discussion}
This work characterizes heterogeneity in decision-making scenarios, by using decision theory to analyze the dimensions on which decision problems can vary.
Our analysis demonstrates how using decision theory to account for context variation in this way can be useful for thinking about generalization in visualization research.
Here, we discuss how decision theory might contribute toward a more robust logic of generalization by offering an alternative to the assumptions of optimistic atomicity and cautious indeterminacy.

\textbf{Utility plays a pivotal role in formalizing decision context.} 
Not only do utility functions $u$ define the desirability of possible actions $a \in A$, but they also provide logical criteria about whether a decision aid $\sigma(s|\omega)$ communicates what it needs to, insofar as utility $u$ defines what task-relevant states $\omega \in \Omega$ are the target for interpretation $\pi(\omega|s)$.
We show that utility also provides a way of operationalizing truthfulness and deception in persuasion~\cite{Kamenica2011-BayesPersuasion} in terms of the alignment of utility for information providers $u_{P}$ and decision-makers $u_{DM}$ on specific task-relevant states $\omega$, and how well a visualization $\sigma$ represents these states.
On this account, the optimistic atomicity view errs in assuming that interpretation and utility are cleanly separable (e.g.,~\cite{Kale2021,Oral2024-decoupling}), such that we can generalize about the former while ignoring the particulars of the later.
Conversely, cautious indeterminacy errs in assuming that describing decision context in an open-ended fashion without formally defining dimensions such as utility will provide a sufficient basis for reasoning about the conditions for generalization.

\textbf{Decision theory provides a deductive framework for a more robust logic of generalization.}
Because decision-making is not a single kind of task, such as we might lump together in a taxonomy (e.g.,~\cite{Amar2005-tasks}), but rather a collection of activities around informed and incentivized action~\cite{brumar2024-typology}, we need a framework in which to accumulate visualization design knowledge that enables us to reason about when certain qualities are likely to be desirable in a decision aid.
We show how decision theory can be useful for conceptualizing variations in decision scenarios, and we envision that similar analyses can be used to \textit{classify decision problems with shared characteristics}. 
For example, decision theory could be used as a deductive framework for qualitative analyses of decision problems, especially in the cast, discover, design, reflect, and write stages of design study methodology~\cite{Sedlmair2012-trenches}, in line with previous extensions of this approach (e.g.,~\cite{Crisan2016-constraints}).
This would provide a systematization of deep knowledge about domain- and context-specific challenges for decision-making, which could then proactively support subsequent efforts to collate findings through research synthesis (cf. ~\cite{Franconeri2021-whatworks, Quadri2022-survey}) and help to spur organized networks of studies on decision-making.

\textbf{Utility elicitation is critically important for visualization in decision-making scenarios.}
Both our analysis and related work by Hullman et al.~\cite{Hullman2011-difficulties} suggest that in settings where the utility of decision-makers and other stakeholders is ambiguous, we need to study their situated values in order to understand what a visualization should communicate.
For example, Wesslen et al.~\cite{Wesslen2022-myopic} describe how their study on how visualizations support investment decisions would be more realistic if it accounted for individual financial goals.
One promising avenue for operationalizing these values as formal utility functions is the use and development of utility elicitation techniques, which are commonplace in behavioral economics but have been explored by relatively little work to date in visualization (e.g.,~\cite{Dimara2018-evaluating, Kale2025-PB, Verma2023-allocation}).
This will be an important direction for future work on tools for decision support, with broad implications for addressing problems around personalization and alignment.

\textbf{Accounting for context in decision problems requires us to recalibrate our expectations during peer review.}
The epistemic commitments of optimistic atomicity and cautious indeterminacy are reinforced aggressively during peer review, and to a lesser extent in the subsequent interpretation of each other's research.
Consider the practice of demanding that discussion sections must include a clear list of generalizable design recommendations (often in bold) as a condition for publication.
This practice, implicitly spurred by the optimistic atomicity assumption, 
asks authors to either overclaim about generalizability or risk violating the reviewer’s expectation that study findings will transcend context.
When reporting on design studies, Sedlmair et al.~\cite{Sedlmair2012-trenches} suggest, ``Design study papers should include only the bare minimum of domain knowledge that is required to understand these abstractions,'' acknowledging that the problem descriptions offered by a such studies may not always be deemed admissible as design knowledge relevant to visualization research.
We argue that we should embrace context-dependence in decision problems and attempt to account for it more systematically in our routines of knowledge construction, rather than reinforcing an inadequate logic of generalization by social convention.

% ...

\acknowledgments{Thanks to Gordon Kindlmann, Jessica Hullman, Pedro Lopes, Krisha Mehta, and Danni Liu for feedback. Special thanks to Gordon for Figure 1. We thank IES \#R305D240023 for funding this work.}


\bibliographystyle{abbrv-doi}

\bibliography{decision-problems}
\end{document}